\documentclass[letter]{jpsj2}

\title{%
A possible ground state and its electronic structure of 
a mother material (LaOFeAs) of new superconductors
}

\author{Shoji Ishibashi$^1$, Kiyoyuki Terakura$^{1,2,3}$ and 
Hideo Hosono$^{4}$} 

\inst{$^1$Research Institute for Computational Sciences (RICS), 
National Institute of Advanced Industrial Science and Technology (AIST),  
1-1-1 Umezono, Tsukuba, Ibaraki 305-8568\\
$^2$Research Center for Integrated Science (RCIS),
Japan Advanced Institute of Science and Technology (JAIST), 
1-1 Asahidai, Nomi, Ishikawa 923-1292\\
$^3$Creative Research Initiative 'Sousei' (CRIS), 
Hokkaido University, Sapporo 001-0021\\
$^4$Frontier Research Center, Tokyo Institute of Technology, 
4259 Nagatsuta, Midori-ku, Yokohama 226-8503}

\abst{The electronic and magnetic properties of the mother material LaOFeAs of new superconductors 
have been carefully studied using 
first-principles 
electronic structure calculations 
based on the generalized gradient approximation in the density functional theory.  
The present calculation predicts that the ground state of LaOFeAs is antiferromagnetic 
with a stripe type magnetic moment alignment leading to orthorhombic symmetry of the crystal.  
In this particular magnetic state, the density of states at the Fermi level is very small. 
On the other hand, LaOFeP has turned out to be paramagnetic and a good metal.  
Implications of the results regarding the experimental observations will also be presented.
}

\begin{document}
\maketitle

\bigskip
\bigskip

\begin{center}
* This letter will be published in J. Phys. Soc. Jpn. {\bf 77} (2008) No.5.
\end{center}
\bigskip
\bigskip

Discovery of a new superconductor always attracts strong attention of the scientific community particularly 
if the critical temperature $T_c$ is above 20 K. 
 It is rather surprising that immediately after the recent news of superconductivity of 
LaO$_{1-x}$F$_x$FeAs\cite{kamihara2008}  several papers, both theoretical and experimental, 
have been circulated in the community. 
There may be four main reasons why this discovery has produced such a strong impact.  
First, $T_c$ is relatively high 26 K.  
Second, it is a new finding that Fe, which is a typical magnetic element, 
seems to be participating to superconductivity.  
Third, the crystal takes again a layered structure and the doping region 
and the superconducting region are geometrically and electronically separated.  
Fourth, the material seems to have strong flexibility 
in the choice of constituent elements suggesting possibility of higher $T_c$ materials 
in the same category.\cite{chen43K}

In order to go further in exploring possibility of higher $T_c$ materials, 
it is essential to understand the basic electronic properties of the mother material LaOFeAs 
and the role of partial replacement of O with F.  
In the analogy of high $T_c$ cuprates,\cite{orenstein} important questions about LaOFeAs 
may be whether magnetism is involved or not and whether it is metallic or insulating.  
These questions have already been addressed by other works\cite{singh,xu,haule} 
in which standard DFT band calculations have been performed. 
At least within our knowledge, it was concluded that the system is metallic and nonmagnetic 
with possible strong antiferromagnetic (AFM) fluctuation. 
We have also performed similar calculations and have arrived at a conclusion 
that the ground state will be a particular AFM state.  
The density of states (DOS) at the Fermi level is very low suggesting that the system may be a bad metal.

Briefly, our calculation is based on the PAW method\cite{blochl} 
and the PBE version\cite {perdew} 
of generalized gradient approximation in the density functional theory. 
Spin-orbit interaction was not taken into account.
We used our in-house computer code named QMAS 
(Quantum MAterials Simulator) 
and have tested the reliability of the code in several ways.  
In relation to the present system, we have performed test calculations 
for LaOMnAs, LaONiP and MnAs with NiAs structure. 
In all of these test calculations, we have confirmed consistency 
between our results and other existing ones\cite{xu,zhao} or experimental facts.\cite{watanabe}  
In the actual calculations for LaOFeAs, we have studied six different magnetic states: 
1) nonmagnetic (NM), 2) ferromagnetic (FM), 3) AFM with G-type (AFM-G, Fig.1a), 
4) AFM with S-type (AFM-S, Fig.1b), 5) AFM with CE1 Type (AFM-CE1, Fig.1c) 
and 6) AFM with CE2 type (AFM-CE2, Fig.1d).  

The convergence criteria were 10$^{-8}$ electrons/bohr$^3$ for the self-consistent-charge distance, 
5$\times$10$^{-5}$ hartree/bohr for the maximum force amplitude 
and 5$\times$10$^{-7}$ hartree/bohr$^3$ for the sum of diagonal components of the stress tensor, 
respectively. The plane-wave energy cutoff was set to 20 hartree. 
The number of {\boldmath $k$} points was set to 12$\times$12$\times$6 for the 1$\times$1 lattice 
(NM, FM and AFM-G), 
10$\times$10$\times$6 for the c(2$\times$2) lattice (AFM-S), 
6$\times$12$\times$6 for the 2$\times$1 lattice (AFM-CE1), or 6$\times$6$\times$6 
for the 2$\times$2 lattice (AFM-CE2).

In order to make a guess which magnetic state is the most stable 
among the above-mentioned six candidates, 
we first estimated the total energy of each magnetic state using the common crystal structure 
determined experimentally.\cite{crystal} 
The results are shown in the 2nd column of Table I 
as the difference from the total energy for the corresponding NM state $\Delta E = E_{(A)FM} - E_{NM}$, 
where $E_{(A)FM}$ and $E_{NM}$ are total energies for (A)FM and NM states.  
The magnetic moment of Fe estimated within the radius of 2.2 bohr is shown in the 3rd column.  
Thus, the AFM-S state is the most stable among the above six candidates 
(NM, FM, AFM-G, AFM-S, AFM-CE1 and AFM-CE2).  
Knowing the most stable magnetic state at this level, 
we optimized the crystal structure (lattice constants and internal coordinates) 
for NM and AFM-S states to evaluate the AFM stability energy more precisely.  
We also optimized the crystal structure for AFM-G because of the reasons mentioned below.  
The revised total energy differences and magnetic moments after structural optimization 
are shown in the 4th and 5th columns, respectively, of Table I. The magnetic stability
energy for both AFM-G and AFM-S is reduced because the effect of structural optimization
is more significant for the NM state. 
In the following, we will discuss basic features of the electronic structures of some selected magnetic states.

Figure 2 shows the density of states (DOS) for the NM state of 
LaOFeAs with the experimental structure,\cite{crystal} 
which is very similar to the corresponding one in other works.\cite{singh,xu} 
The whole valence states are divided into p bands coming from O and As and d bands coming from Fe.  
 
The width of d band is about 4 eV, being about 2/3 of that of metallic bcc Fe and suggesting significant itinerant character of d states.  
By counting the capacity of bands and number of valence electrons, 
the formal number of d electrons in LaOFeAs is six per formula unit 
and the Fermi level is located at the shoulder just below the dip in the d band.  
We have found that there is only very little contribution 
from oxygen p states to the states at the Fermi level.  
As is well known in the physics of transition metals,\cite{terakura} 
this situation of the Fermi level with regard to the d band may favor AFM instability 
rather than FM one if any magnetic instability may exist.  
This is indeed the case as Table I shows.  
However, it is fairly difficult to find the most stable AFM states 
because there are infinite number of possible metastable AFM states.  
In the present system, as Fe forms a square lattice, 
the AFM-G state may naively be expected to be realized.  
With the experimental crystal structure, the AFM-G state is more stable 
than the NM state by 0.08 eV per formula unit. (Note that the AFM-G
stability energy with respect to the NM state is reduced to 0.016 eV after
structural optimizatin. See TableI.)
However, this AFM state has a rather strange feature in DOS, which is shown in Fig.3.  
Due to the G-type arrangement of the exchange potential, 
d states will have a tendency of localization. 
This tendency is particularly strong for d$_{3z^2-r^2}$ due to both the 2 dimensional network 
and the flattening of the tetrahedral coordination of As. 
This strong localization of d$_{3z^2-r^2}$ produces a rather sharp peak in the region of a dip in DOS 
where the Fermi level is located.  
With such a sharp peak in DOS at the Fermi level, 
we generally expect some structural distortion. 
Although we have tested several possible structural distortion models, 
none of them could remove the sharp peak from the Fermi level. 

Instead of exploring further possibilities of structural distortion within
the AFM-G magnetic structure, we studied three other types of AFM spin arrangement, 
which reduce the symmetry of the crystal from tetragonal (for NM, FM and AFM-G) to orthorhombic. 
These magnetic structures look strange at first sight, 
but the angular variation of d orbitals can adjust the covalency between Fe atoms to be consistent 
with the magnetic structure.  In other words, magnetic structure is accompanied 
with orbital ordering and then crystal structure is consistently modified 
from the tetragonal symmetry. Such examples are frequently observed 
in perovskite transition metal oxides.\cite{tokura} 
Our calculation predicts that the AFM-S state is the most stable among six possible magnetic states studied. 
We therefore present some more details of the optimized crystal structure 
and the electronic structure of the AFM-S state below.
 
As mentioned above, the orthorhombic symmetry of the magnetic moment arrangement in the AFM-S state 
will lead to modification of the crystal structure mediated by orbital ordering, 
which can be seen in the spin density distribution in Fig.4. 
Qualitatively, the states of local majority spin state with energy 
near the Fermi level contribute to the spin density.  
As the states in this energy range are more or less antibonding between nearest neighbor Fe atoms, 
the spin density is more localized along the stripe of FM spin alignment. 
According to the structural optimization, the lattice constants of the orthorhombic cell, 
which is originally the c(2$\times$2) cell ($a = 4.03533 \times \sqrt{2} = 5.70682$ \AA, $c = 8.74090$\ \AA) of the tetragonal lattice,  
are $a = 5.686$\ \AA\ (parallel to the stripe), $b = 5.758$\ \AA\ (perpendicular to the stripe) and $c = 8.707$\ \AA. The $z$ values for 
La and As ($z_{La}$ and $z_{As}$) are 0.142 and 0.646, respectively.
For comparison, the optimized tetragonal lattice constants for the NM state are 
$a = 5.677$ \AA, $c = 8.640$\ \AA\ with $z_{La} = 0.145$ and $z_{As} = 0.639$.
Figure 5 shows the total DOS and Fe d partial DOS of the AFM-S state. 
Orbital decomposed partial DOSs in the AFM-S state are shown in Fig.6
The band dispersion in the Brillouin zone of the c(2$\times$2) unit cell of the original lattice is shown in Fig.7. 
Although a couple of bands cross the Fermi level, very small DOS at the Fermi level 
(0.246 states/eV/cell for each spin) implies 
that the system may be a bad metal.  The states at the Fermi level have contributions dominantly 
from minority-spin d$_{yz}$ and d$_{3z^2-r^2}$ orbitals and less dominantly 
from majority-spin d$_{yz}$ and d$_{x^2-y^2}$ orbitals. 
The contribution from As p$_y$ is noticeable but an order of magnitude smaller 
than the contribution from Fe d orbitals. Note that the $x$ ($y$) axis is parallel (perpendicular) to the stripe 
of the AFM-S structure. 

As for the electronic structure of the NM LaOFeP, we obtained very similar results to 
those reported by Kamihara {\em et al.}\cite{kamihara2006} 
Now we ask ourselves how different it is between LaOFeAs and LaOFeP 
in our band structure calculations.  
Their NM states have nearly identical DOS except 
that the band width is slightly narrower in LaOFeAs than in LaOFeP. There is also some
noticeable difference in the contributions from As and P to the states at the Fermi level:
p state contribution from As is much larger than that from P. 
Such a subtle difference in the NM state produces striking difference between them in the magnetic state.  
We have found that the AFM-S state converges to the NM state very slowly 
in the process of structural optimization for LaOFeP.  
Therefore, LaOFeP is paramagnetic and a good metal with possible AFM fluctuation corresponding to the S type.

The results of the present calculation are qualitatively consistent 
with some experimental data such as resistivity.\cite{kamihara2008,kamihara2006} 
As was already mentioned above, the calculated electronic structure implies that LaOFeAs 
may be a bad metal while LaOFeP may be a good metal. 
Qualitatively such behavior is actually observed experimentally.\cite{kamihara2008,kamihara2006} 
Stability of AFM state and very small DOS at the Fermi level for LaOFeAs may be consistent 
with the fact that the undoped pure system does not show superconductivity. 
Note that pure LaOFeP is a superconductor below 4 - 7 K.\cite{kamihara2006} 
On the other hand, we do not know at the present stage 
how to explain the observed magnetic susceptibility of LaOFeAs\cite{kamihara2008} in terms of AFM ordering 
and small DOS at the Fermi level.  Experimentally, the susceptibility is almost temperature independent 
like a metallic Pauli susceptibility and the magnitude of the order of about 0.5 x $10^{-3}$ emu/mol 
is too large to be explained by DOS at the Fermi level. 
The role of partial oxygen replacement with F is now being studied.

We thank Prof. H. Fukuyama for useful discussion and Dr. M. Kohyama, Dr. S. Tanaka and Dr. T. Tamura 
for their help in developing our computational code QMAS. 
The present work is partially supported by the Next Generation Supercomputer Project, 
Nanoscience Program and also partly by Grant-in-Aids 
for Scientific Research in Priority Area ^^ ^^ Anomalous Quantum Materials'', both from MEXT, Japan. 
The calculations were performed using the AIST Super Cluster at the Tsukuba Advanced Computing
Center (TACC), AIST.

\begin{figure}[ht]
\caption{Antiferromagnetic order patterns investigated in this study.
a: AFM-G, b: AFM-S, c: AFM-CE1, d: AFM-CE2. 
Squares and a rectangular in broken lines represent unit cells for the corresponding AFM states.
Schematic view of the experimetal unit cell is shown also.}
\label{f1}
\end{figure}

\begin{figure}[ht]
\caption{Density of states (DOS) for the NM state of 
LaOFeAs with the experimental structure. 
The energy zero corresponds to the Fermi level.}
\label{f2}
\end{figure}

\begin{figure}[ht]
\caption{Density of states (DOS) for the AFM-G state of 
LaOFeAs with the experimental structure. 
The energy zero corresponds to the Fermi level.
The upper half represents the DOS for up spin while the lower half is for down spin.}
\label{f3}
\end{figure}

\begin{figure}[ht]
\caption{Spin density distribution viewed along the $c$ direction 
for the AFM-S state of LaOFeAs.
Isodensity surface contours for the difference 
between the up-spin and down-spin electron densities are plotted. 
The blue and red surfaces correspond to $\pm$ 0.005 electrons/bohr$^3$, respectively.
Larger surfaces are located on Fe atoms. As atoms are at ($x$,$y$) = 
(0.25,0.25), (0.75,0.25), (0.25,0.75) and (0.75,0.75).
The boundary corresponds to the square shown in Fig.1b.
Along {\boldmath $a$} ({\boldmath $b$}), the lattice shrinks (expands) 
from the original square.}
\label{f4}
\end{figure}

\begin{figure}[ht]
\caption{Total DOS (a) and Fe d partial DOS (b) 
for the AFM-S state of LaOFeAs. The energy zero corresponds to the Fermi level.
The upper half represents the DOS for up spin while the lower half is for down spin.}
\label{f5}
\end{figure}

\begin{figure}[ht]
\caption{Orbital decomposed partial DOS 
for the AFM-S state of LaOFeAs. The energy zero corresponds to the Fermi level.
The upper half represents the DOS for up spin while the lower half is for down spin.}
\label{f6}
\end{figure}

\begin{figure}[ht]
\caption{Band dispersion in the Brillouin zone of 
the c(2$\times$2) unit cell of the original lattice 
for the AFM-S state of LaOFeAs.
The right panel is a magnified drawing of the left one
near the Fermi level. The energy zero corresponds to the Fermi level.
Wave vector is shown in the unit of ($\pi/a$,$\pi/b$,$\pi/c$).}
\label{f7}
\end{figure}

\begin{table*}
\caption{2nd column: Total energy of each magnetic state with reference to that of NM state
with the common lattice structure from experiment\cite{crystal}, 3rd column: Magnetic 
moment of Fe, 4th column: the same quantity as that of column 2 after structural optimization
only for AFM-G and AFM-S, 5th column: Magnetic moment of Fe after structural optimization.
\label{table1}}
\bigskip

\begin{center}
\begin{tabular}{lrrrr} \hline
 
 Type & $\Delta E$$^*$ (eV/formula) & M$^{**}$ ($\mu_B$) 
& $\Delta E_{opt}$$^*$ (eV/formula) & M$_{opt}^{**}$ ($\mu_B$) \\ \hline
 FM & $- 0.004$ & 0.40 & - & - \\ 
 AFM-G & $- 0.080$ & 1.89 & $- 0.016$ & 1.51 \\ 
 AFM-S & $- 0.174$ & 2.12 & $- 0.098$ & 2.00 \\
 AFM-CE1 & $- 0.064$ & 1.88 & - & - \\
 AFM-CE2 & $- 0.116$ & 2.12 & - & - \\ 
\hline
\end{tabular}
\medskip

$^*$ $\Delta E = E_{(A)FM} - E_{NM}$\\ 
$^{**}$ Magnetic moments were estimated within a radius of 2.2 bohr.
\end{center}

\end{table*}

\end{document}